\documentclass[prd,twocolumn,preprintnumbers,amsmath,amssymb,superscriptaddress,showpacs,showkeys,nofootinbib]{revtex4-1}
\usepackage{color}
\usepackage{amsmath,amsfonts,amssymb,amsthm,amstext,amscd,eucal,srcltx,mathrsfs}
\usepackage{epsfig,graphicx,bm}
\usepackage{epstopdf, epsf}
\usepackage{dcolumn} 
\usepackage{hyperref}
\newcommand{\be}{\begin{eqnarray}}
\newcommand{\ee}{\end{eqnarray}}
\newcommand{\bra}[1]{\mbox{$\langle\, #1 \mid$}}

\newcommand{\ket}[1]{\mbox{$\mid #1\,\rangle$}}

\renewcommand{\d}{\mbox{${\rm d}$}} 
\newcommand{\lp}{\ell_{\rm p}}
\newcommand{\mpl}{m_{\rm p}}
\newcommand{\gn}{G_{\rm N}}
\newcommand{\rh}{r_{\rm H}}
\newcommand{\Rh}{R_{\rm H}}

\newcommand{\Rl}{R_{\rm L}}
\newcommand{\psis}{{\psi}_{\rm S}}

\newcommand{\psih}{{\psi}_{\rm H}}

\begin{document}
\title{Quantum Formation of Primordial Black holes}
\author{Roberto Casadio}
%
\affiliation{Dipartimento di Fisica e Astronomia,
Alma Mater Universit\`a di Bologna,
via~Irnerio~46, 40126~Bologna, Italy}
\affiliation{I.N.F.N., Sezione di Bologna, IS FLAG
viale~B.~Pichat~6/2, I-40127 Bologna, Italy}
\author{Andrea Giugno}
\affiliation{Arnold Sommerfeld Center, Ludwig-Maximilians-Universit\"at,
Theresienstra{\ss}e 37, 80333 M\"unchen, Germany}
\email{A.Giugno@physik.uni-muenchen.de}
\author{Andrea Giusti}
\affiliation{Dipartimento di Fisica e Astronomia,
Alma Mater Universit\`a di Bologna,
via~Irnerio~46, 40126~Bologna, Italy}
\affiliation{I.N.F.N., Sezione di Bologna, IS FLAG
viale~B.~Pichat~6/2, I-40127 Bologna, Italy}
\affiliation{Arnold Sommerfeld Center, Ludwig-Maximilians-Universit\"at,
Theresienstra{\ss}e 37, 80333 M\"unchen, Germany}
\author{Michele Lenzi}
\affiliation{Dipartimento di Fisica e Astronomia,
Alma Mater Universit\`a di Bologna,
via~Irnerio~46, 40126~Bologna, Italy}
\affiliation{I.N.F.N., Sezione di Bologna, IS FLAG
viale~B.~Pichat~6/2, I-40127 Bologna, Italy}
\begin{abstract}
We provide a (simplified) quantum description of primordial black holes at the time of
their formation.
Specifically, we employ the horizon quantum mechanics to compute the probability of black hole
formation starting from a simple quantum mechanical characterization of primordial density fluctuations
given by a Planckian spectrum.
We then estimate the initial number of primordial black holes in the early universe as a function of their
typical mass, spatial width and temperature of the fluctuation.
\end{abstract}
%
%
%
%
\maketitle
\section{Introduction}
The interest in primordial black holes (PBH) started over fifty years ago~\cite{zeldovich,hawking},
and the conjecture was soon put forward that they could account for a (possibly significant) fraction of
the dark matter~\cite{carrH}.
The basic idea is that, in the early radiation dominated Universe, a sufficiently overdense region 
should collapse into a black hole~\cite{carr}.
Many mechanisms to generate primordial fluctuations of sufficient density have then been proposed and
their confrontation with astrophysical and cosmological data has generated a huge literature (for a review,
see, e.g.~Ref.~\cite{sasaki}).
\par
In this work, we are interested in the fundamental issue of the formation of PBH's.
In fact, the importance of the spatial profile of perturbations in classical general relativity has already
been pointed out in Refs.~\cite{germani,musco,garriga}.
Our aim here is to show that the quantum nature of primordial fluctuations and the overall process of
black hole formation could also be very relevant.
For this purpose, we shall consider a simplified scenario in which we can carry out a complete
analysis, albeit without the presumption to obtain predictions directly comparable with the 
observations.
For estimating the initial number of PBH's, we shall then employ the Horizon Quantum Mechanics
(HQM)~\cite{fuzzyh,hqft}, which was precisely proposed with the purpose of describing the
gravitational radius of spherically symmetric compact sources and determining the existence
of a horizon in a quantum mechanical fashion.
\section{Quantum primordial black holes}
We shall here model a primordial fluctuation as a quantum state of excited gravitons
with a thermal distribution above the de~Sitter ground state~\cite{dvali},  and then employ
the (global) HQM~\cite{fuzzyh,hqft} in order to compute the probability that this fluctuation
is a black hole.
\par
The corpuscular picture of gravity~\cite{dvali} was first introduced for describing black holes,
but it also applies to cosmology and inflation in particular~\cite{CK,inflation}.
In order to have the de~Sitter space-time in general relativity, one must assume the existence of a
cosmological constant $\Lambda$, or vacuum energy density $\rho_L$, so that the Friedman equation
reads~\footnote{We shall use units with $c=k_{\rm B}=1$, $\gn=\lp/\mpl$ and $\hbar=\lp\,\mpl$.}
\be
H^2
\equiv
\left(\frac{\dot a}{a}\right)^2
=
\gn \, \rho_L
\ .
\ee
Upon integrating on the volume inside the Hubble radius
\be
L = H^{-1} = \sqrt{{3}/{\Lambda}}
\ ,
\ee
we obtain
\be
L
\simeq
\gn\,L^3\,\rho_L
\simeq
\gn\,E_L
\ .
\label{LgM}
\ee
The length $L$ therefore satisfies a relation exactly like the Schwarzschild radius for a black hole of ADM mass
$E_L$, which supports the conjecture that the de~Sitter space-time could likewise be described
as a condensate~\cite{inflation}.
One can roughly describe the corpuscular model by assuming that the graviton self-interaction gives rise
to a condensate of $N$ (soft virtual) gravitons of typical Compton length of the order of $L$, so that 
$
E_L \simeq N\,{\lp\,\mpl}/{L}
$
and the usual consistency conditions 
\be
N\simeq E_\Lambda^2/\mpl^2
\ee
turns out to be a natural consequence~\cite{dvali}.
Equivalently, one finds
\be
L \simeq \sqrt{N}\,\lp
\ ,
\ee
which shows that for a macroscopic black hole, or universe,
one needs $N\gg 1$.
\subsection{Schwarzschild-de~Sitter space-time}
For our analysis of primordial perturbations, we shall employ the spherically symmetric and static
Schwarzschild-de~Sitter metric 
\be
\d s^2
=
-f \, \d t^2
+f^{-1}\,\d r^2
+r^2 \,\d \Omega^2 
\ ,
\label{SdS}
\ee
with
\be
f=1 - \frac{2 \, \gn \, M}{r} - \frac{r^2}{L^2}
\ .
\ee
This metric represents the exterior of a black hole of mass $M$ as seen by a static observer located at constant
radial position $r$, provided $\Rh<r<\Rl$, where $\Rh$ is the black hole horizon and $\Rl$ the cosmological horizon.
For our purpose, we can associate $L$ with the background homogenous space-time undergoing inflation,
and the mass $M$ with the energy of the local fluctuation.
\par
As usual, horizons are given by real solutions of the equation $f(r)=0$, that is
\be 
R_{\rm H/L}
=
\frac{2\,L}{\sqrt{3}}\,
\cos \left[ \frac{\pi}{3} \pm \frac{1}{3}\,\arccos\left(\frac{3\,\sqrt{3}\, \gn \, M}{L}\right) \right]
\ ,
\label{r-hor}
\ee
and the nomenclature is then justified by the fact that $R_{\rm H} < R_{\rm L}$ for 
$L\ge 3\,\sqrt{3}\,\gn\,M\ge 0$, with the proper metric signature $(-+++)$ in the region
$R_{\rm H}< r< R_{\rm L}$, as anticipated above.
Moreover, for $L\gg \gn\,M$, the black hole horizon approaches the usual Schwarzschild radius,
that is
\be
R_{\rm H}  
\simeq
2 \,\gn\, M 
\left[ 1 + \left(\frac{R_M}{L}\right)^2 \right] 
\ .
\label{r-hor-asym}
\ee
In the same limit, the cosmological horizon approaches the de~Sitter value,
\be
R_{\rm L} 
\simeq
L
\left(1 - \frac{R_{\rm M}}{2\,L}\right) 
\ . 
\label{r-L-asym}
\ee
Finally, we note that the extremal configuration is
\be 
R_{\rm H} =R_{\rm L} = 3\,\gn\,M = {L}/{\sqrt{3}}
\ ,
\label{extremal}
\ee
in which case the coordinate $r$ is always time-like and there are no values
of $r$ corresponding to a static observer.
Nonetheless, since $2\,\gn\,M\le \Rh(M,L)\le 3\,\gn\,M$, in the following we shall
consider the case $\Rh\simeq 3\,\gn\,M$ for simplicity and for maximising the probability
of black hole formation.
\subsection{Horizon Quantum Mechanics}
According to this approach~\cite{fuzzyh,hqft}, we assume the existence of two observables,
the quantum Hamiltonian corresponding to the energy $M$ of the fluctuation which
might result in a black hole,
\be
\hat H
=
\sum_\alpha
E_\alpha\ket{E_\alpha}\bra{E_\alpha}
\ ,
\label{HM}
\ee
and the gravitational radius corresponding to the black hole horizon, with eigenstates
\be
\hat r_{\rm H}\,\ket{{\rh}_\beta}
=
{\rh}_\beta\,\ket{{\rh}_\beta}
\ .
\ee
The cosmological length $L$, being associated with the background space-time,
is instead regarded as a classical parameter here, like the electric charge
of the Reissner-Nordstr\" om space-time in Refs.~\cite{RN}~\footnote{A more general
treatment in which $L$ is also quantised is left for future developments (see also Ref.~\cite{CK}).}. 
\par
General quantum states for the fluctuation can be described by linear combinations of 
the form
\be
\ket{\Psi}
=
\sum_{\alpha,\beta}
C(E_\alpha,{\rh}_\beta)\,
\ket{E_\alpha}
\ket{{\rh}_\beta} 
\ ,
\label{Erh}
\ee
but only those for which the relation~\eqref{r-hor} between the Hamiltonian and the
gravitational radii hold are viewed as physical.
In particular, we invert Eq.~\eqref{r-hor-asym} in order to write
\be
M(\Rh;L)
=
\frac{\Rh}{2\,\gn}
\left(1-\frac{\Rh^2}{L^2}\right)
\ ,
\ee
and then impose this condition as the weak Gupta-Bleuler constraint
\be
0
&=&
\left[
\hat H
-
M(\hat r_{\rm H};L)
\right]
\ket{\Psi}\\
&=&
\sum_{\alpha,\beta}
\left[
E_\alpha
-
M({\rh}_\beta;L)
\right]
C(E_\alpha,{\rh}_\beta)\,
\ket{E_\alpha}
\ket{{\rh}_\beta} \notag
\ .
\label{Hcond}
\ee
The solution is given by
\be
C(E_\alpha,{\rh}_\beta)
=
C\left(E_\alpha,\Rh(E_\alpha;L)\right)
\delta_{\alpha\beta} \ ,
\ee
which means that Hamiltonian eigenmodes and gravitational radius eigenmodes can only appear
suitably paired in a physical state.
\par
By tracing out the gravitational radius, we recover the spectral decomposition of the system,
\be
\ket{\psis}
&=&
\sum_\alpha
C\left(E_\alpha,\Rh(E_\alpha,L)\right)
\ket{E_\alpha}
\nonumber
\\
&\equiv&
\sum_\alpha
C_{\rm S}(E_\alpha,L)\,\ket{E_\alpha}
\ ,
\ee
in which we used the (generalised) orthonormality of the gravitational radius eigenmodes~\cite{hqft}.
Conversely, by integrating out the energy eigenstates, we obtain the Horizon Wave-Function (HWF)~\cite{fuzzyh}
\be
\psih({\rh}_\alpha)
=
C_{\rm S}(M({\rh}_\alpha,L))
\ .
\label{psihd}
\ee
If the index $\alpha$ is continuous (again, see Ref.~\cite{hqft} for some important remarks),
the probability density that we detect a
gravitational radius of size $\rh$ associated with the quantum state $\psis$ is given by
\be
\mathcal{P}_{\rm H}(\rh)
=
4\,\pi\,\rh^2\,|\psih(\rh)|^2
\ ,
\ee
and we can define the conditional probability density that the source lies
inside its own gravitational radius $\rh$ as
\be
\mathcal{P}_<(r<\rh)
=
P_{\rm S}(r<\rh)\,\mathcal{P}_{\rm H}(\rh)
\ ,
\label{PrlessH}
\ee
where
\be
P_{\rm S}(r<\rh)= 4\,\pi \int_0^{\rh} |\psis(r)|^2\,r^2\,\d r
\ .
\ee
Finally, the probability that the system in the state $\psis$ is a black hole
will be obtained by integrating~\eqref{PrlessH} over all possible values of $\rh$,
namely
\be
P_{\rm BH}
=
\int_0^\infty
\mathcal{P}_<(r<\rh)\,\d \rh
\ .
\label{pbhgen}
\ee
\subsection{Thermal density fluctuations in de~Sitter}
As we mentioned above, we assume the spectral decomposition of a primordial perturbation
is given by a Planckian distribution at the temperature $T=k\,T_{\rm dS}$, that is
\be
C_{\rm S}^2(E)
\simeq
\frac{\mathcal{N}^2}{T^{3}}\,
\frac{(E-E_L)^2}
{{\exp\left\{(E-E_L)/T\right\}-1}}
\ ,
\label{C(E)ThermApprox}
\ee
where $E_L$ is the background de~Sitter energy in Eq.~\eqref{LgM} and the de~Sitter
temperature $T_{\rm dS}\simeq \mpl\,\lp/L$.
The mean energy density above the ground state associated to such a fluctuation is thus
given by
\be
\frac{\Delta E}{E_L}
&\simeq&
\int_{E_L}^\infty
\frac{(E-E_L)}{E_L}\,C_{\rm S}^2(E)\,\d E
\nonumber
\\
&\simeq&
\frac{\pi^4\, T}{30\,\zeta(3)\,E_L}
\simeq
2.7\,\frac{k}{N}
\ ,
\label{dE}
\ee
which implies that a fluctuation can carry a significant fraction of the energy within the length $L\sim \sqrt{N}$
only if the temperature is $k\sim N$ times the de~Sitter temperature $T_{\rm dS}\sim 1/\sqrt{N}$.
Let us also note that the above result $\Delta E/E_L\sim 1/N$ for $k=1$ is analogous to the one for
thermal corpuscular black holes~\cite{thbh}.
\subsection{Black hole formation}
From the spectral decomposition of the whole fluctuation~\eqref{C(E)ThermApprox}, on assuming
the extremal relation~\eqref{extremal}, that is 
\be
\rh \simeq 3\,\lp\,\frac{E-E_L}{\mpl}
\ ,
\ee
one immediately finds the HWF
\be
\psih(\rh)
\simeq
\frac{\mathcal{N}_{\rm H}\,(L/k)^{5/2}\,\rh/\lp^{5}}
{\sqrt{\exp\{L\,\rh/3\,k\,\lp^2\}-1}}
\ ,
\ee
with $\mathcal{N}_{\rm H}= 1/108\,\sqrt{2\,\pi\,\zeta(5))}\simeq 3.6\cdot 10^{-3}$ and where we used
$T=k\,T_{\rm dS}=k\,\mpl\,\lp/L$.
\begin{figure}[t]
\centering
\includegraphics[width=8cm]{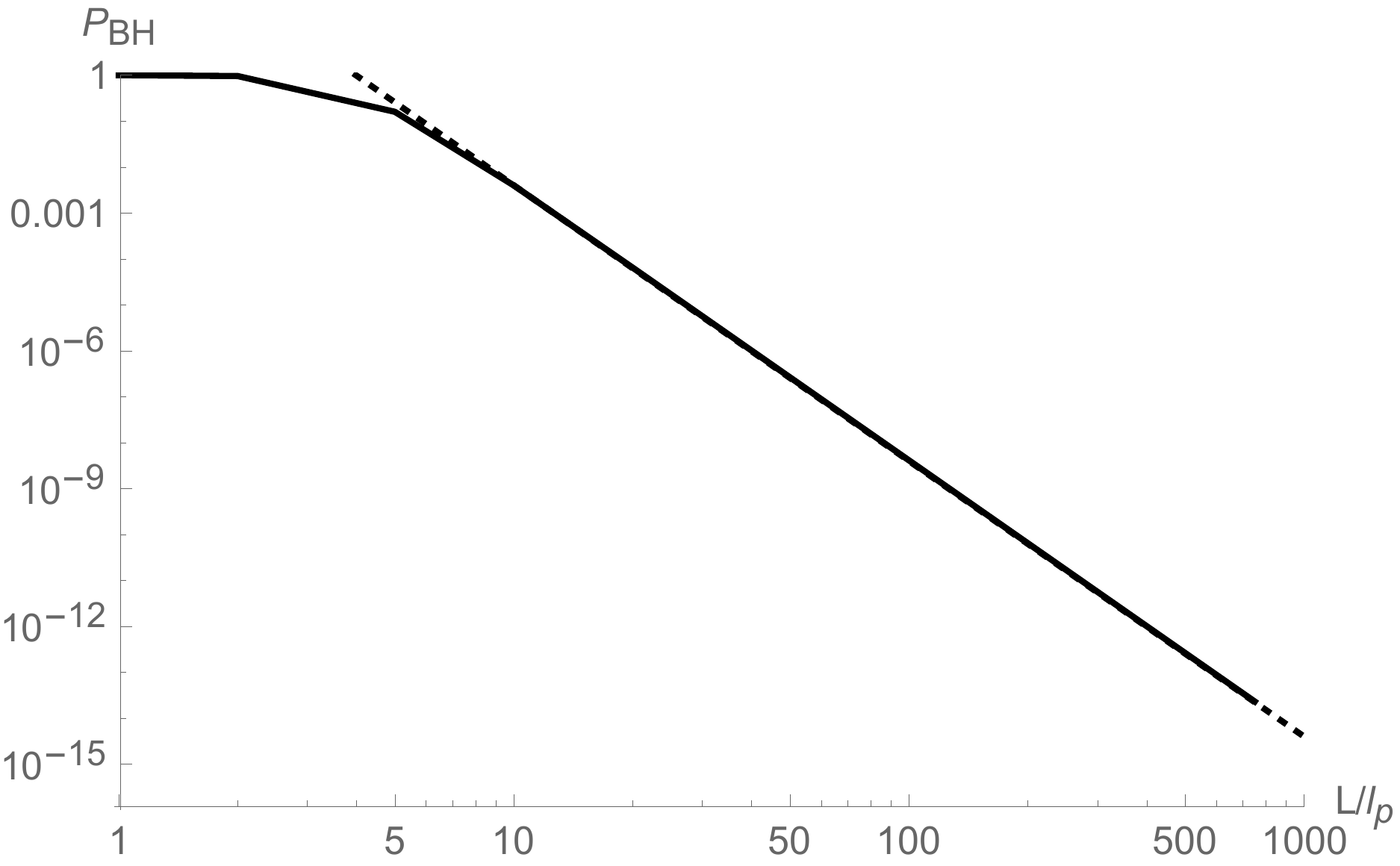}
\caption{Black hole probability for $\lambda=L$ (solid line) vs.~analytical
approximation~\eqref{PbhnnT} (dotted line) with $T=T_{\rm dS}$ (equivalent to $k=1$).
}
\label{pbhnnLl}
\end{figure}
\par
In order to proceed, we describe the fluctuation in position space by means of a Gaussian
wave-function of width $\lambda\sim L$, 
\be
\psi(r)
=
\frac{e^{-\frac{r^2}{2 \, \lambda^2}}}{\left(\lambda \, \sqrt{\pi}\right)^{3/2}}
\ ,
\label{psiSb}
\ee
from which we then obtain 
\be
\notag
\mathcal{P}_<(r<\rh)
&\simeq&
\frac{(L/k)^5\,\rh^4}{5832\,\zeta(5)\,\lp^{10}
\left(e^{\frac{L\,\rh}{3\,k\,\lp^2}}-1\right)^2}
\\
&&
\times
\left[
{\rm Erf}\left(\frac{\rh}{\lambda}\right)
-\frac{2\,e^{-\frac{\rh^2}{\lambda^2}}\,\rh}{\sqrt{\pi}\,\lambda}
\right]
\ .
\label{PlRhk}
\ee
Upon integrating this expression over $\rh$ (numerically) for fixed values of $\lambda$, $L$
and $k=T/T_{\rm dS}$ one obtains the probability~\eqref{pbhgen} that the fluctuation is a black hole.
\par
\begin{figure}[h]
\centering
\includegraphics[width=8cm]{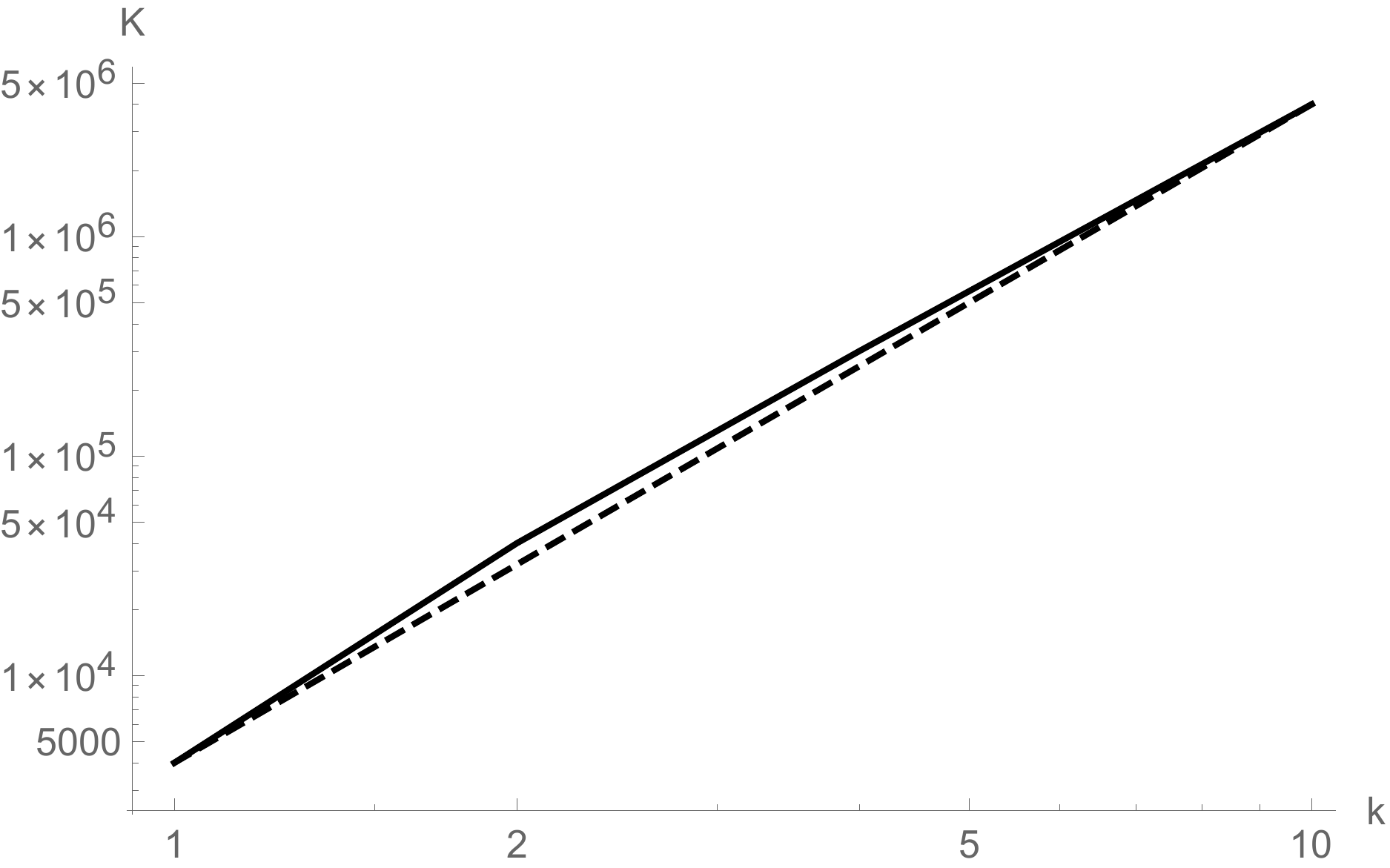}
\caption{Coefficient $K_{\lambda=L,k}$ evaluated numerically (solid line) and analytical
approximation~\eqref{aKk}.
}
\label{Kk}
\end{figure}

The result as a function of $L$ for $k=1$ and $\lambda=L$ is plotted in Fig.~\ref{pbhnnLl}, and,
for $L\gtrsim 10\,\lp$, it is extremely well approximated by
\be
P_{\rm BH}(L)
\simeq
K_{\lambda=L,k=1}\left(\frac{\lp}{L}\right)^6
\ ,
\label{PbhnnT}
\ee
with $K_{L,1}\simeq 4\cdot 10^3$.
For values of $\lambda<L$, the probability $P_{\rm BH}$ remains of the form in Eq.~\eqref{PbhnnT},
with $K_{\lambda=L/2,k=1}\simeq 4\cdot 10^4$ and $K_{\lambda=L/4,k=1}\simeq 3\cdot 10^5$.
For larger values of the temperature (that is, for $k>1$), we notice that the function in front of the
square brackets in Eq.~\eqref{PlRhk} depends on the effective length $L/k$.
We therefore expect that doubling the temperature is (roughly) equivalent to halving the Hubble scale $L$. In general, we find that for $L\gg \lp$, the coefficient in Eq.~\eqref{PbhnnT} is very well approximated by
\be 
K_{\lambda,k}
\simeq
4\cdot 10^3
\left(\frac{k\,L}{\lambda}\right)^3
\ ,
\label{aKk}
\ee
as is shown in Fig.~\ref{Kk}.
Upon including this result, one finally obtains
\be
P_{\rm BH}(L)
\simeq
4\cdot 10^3
\left(\frac{T}{T_{\rm dS}}\right)^3
\left(\frac{\lp}{\lambda}\right)^3
\left(\frac{\lp}{L}\right)^3
\ ,
\label{PbhF}
\ee
which holds for $L\gtrsim \lambda\gg \lp$ and $T\gtrsim T_{\rm dS}$.
\par
Since the typical mass $M$ of these PBH is related to $L$ according to Eq.~\eqref{dE}, that is
$M\simeq E_\Lambda \simeq \mpl\,{L}/{\lp}$, we can equivalently write 
\be
P_{\rm BH}(M)
\simeq
30
\left(\frac{T}{T_{\rm dS}}\right)^3
\left(\frac{\lp}{\lambda}\right)^3
\left(\frac{\mpl}{M}\right)^3
\ ,
\label{PbhnnM}
\ee
for $M\gtrsim \mpl/10$.
We remark that there appears no sharp threshold in the black hole mass,
which is in fact what one expects from the HQM~\cite{fuzzyh}.
Upon multiplying this probability for the number of de~Sitter patches of size $L\sim M$, one
can estimate the total number of PBH's inside the visible universe of a given mass,
\be
N_{\rm BH}(M)
&\simeq&
\frac{R_0\,\mpl}{3 \, \sqrt{3} \, \lp\,M}\,P_{\rm BH}(M)
\nonumber
\\
&\simeq&
6 \cdot 10^{62}
\left(\frac{T}{T_{\rm dS}}\right)^3
\left(\frac{\lp}{\lambda}\right)^3
\left(\frac{\mpl}{M}\right)^4
\ ,
\label{N(M)}
\ee
in which we used $R_0\simeq 10^{62}\,\lp$ for the size of the visible Universe.
We remark that this counting applies to the initial number PBH's and neglects
both the subsequent evaporation and possible accretion.
\par
The above result can be recast in the following form.
First we note that the de~Sitter energy density is holographic~\cite{DEC},
that is
\be
\rho_L
\simeq 
\frac{3\, \mpl}{8\, \pi\, \lp\,L^2}
\ ,
\ee
whereas the energy density of the fluctuation is given by
\be
\delta\rho
\simeq
\frac{\Delta E}{(4\, \pi\, \lambda^3 / 3)}
\simeq
\frac{\pi^3\,T}{40\, \zeta (3)\,{\lambda^3}}
\simeq
0.7 \, \frac{T}{\lambda^3}
\ .
\ee
Upon recalling that $T_{\rm dS}\simeq \mpl\,\lp/L$, one finds
\be
\frac{\delta\rho}{\rho_L}
\simeq
6 \, \frac{T}{T_{\rm dS}}
\left(
\frac{\lp}{L}
\right)^2
\left(
\frac{L}{\lambda}
\right)^3
\ ,
\ee
and therefore
\be
\notag
N_{\rm BH}(M)
&\simeq&
2\cdot 10^{58} \, \left( \frac{\lambda}{\lp} \right)^6 \left( \frac{\delta \rho}{\rho _L} \right)^3 \left( \frac{\mpl}{M} \right)^7 \\
&\simeq&
5 \cdot 10^{62} \left( \frac{\delta \rho}{\rho _L} \right)^3\frac{\mpl}{M}
\ ,
\label{N(M)}
\ee
where we assumed $\lambda\simeq L$ in the last approximation.
\smallskip
\null
\smallskip
\null
\smallskip
\section{Conclusions and outlook}
In this work, we have computed the probability of PBH formation in a simple framework
for the early Universe and quantum density perturbations.
Our results should first and foremost caution that the details of the process of black hole formation still need
to be understood better and that quantum effects might not be negligible.
\par
In particular, after a brief review of the HQM, we have provided an explicit computation of the probability
of black hole formation by describing the primordial fluctuations in terms of a Planckian distribution
of typical temperature $T \simeq k \, T_{\rm dS}$.
The factor $k$ was left arbitrary here, but it should be easy be obtained it from any specific models
in the literature.
\par
The key result of our analysis is that the mass spectrum of PBH's~\eqref{N(M)} appears to be extremely
suppressed in this simplified setup.
In particular, it seems that a purely quantum mechanical treatment of primordial density perturbations
implies a very low likelihood for the formation of PBH's in the very early Universe, unless one has
reasons to consider for the fluctuations some very large values of $\delta \rho / \rho_L\sim T/T_{\rm dS}$.
\par
As we mentioned in the Introduction, much of the present interest stems from PBH's as candidates
for the dark matter~\cite{carrH,Carr:2016drx}.
This of course requires a significant production which could occur during inflation only during stages
of departure from the (quasi de~Sitter) slow-roll evolution, or later on during the radiation dominated
era~\cite{Carr:2018nkm,Cai:2018rqf,Khlopov:2008qy,Niemeyer:1997mt}.
The calculation in the present work should therefore be adapted to such scenarios in order
to say something of relevance in this respect. 
%
%
%
\acknowledgements
The authors are grateful to G.~Dvali for helpful discussions.
R.~C., A.~Giusti and M.~Lenzi are partially supported by the INFN grant FLAG.
A.~Giugno is partially supported by the ERC Advanced Grant
339169 ``Selfcompletion".
\end{document}